\documentclass[traditabstract]{aa}
\usepackage{epsfig,graphicx}
\usepackage{natbib}
\bibpunct{(}{)}{;}{a}{}{,}    

\begin{document}

\title{Horizontal or vertical magnetic fields on the quiet Sun:}
\subtitle{Angular distributions and their height variations}
\titlerunning{Horizontal or vertical magnetic fields  on the quiet Sun}
\author{J.O. Stenflo\inst{1,2}}
\institute{Institute of Astronomy, ETH Zurich, CH-8093 Zurich \and Istituto Ricerche Solari Locarno, Via Patocchi, CH-6605 Locarno Monti, Switzerland}

\date{}

\abstract{
Different analyses of identical Hinode SOT/SP data of quiet-sun
magnetic fields have in the past led to contradictory 
answers to the question of whether the angular distribution of field
vectors is preferentially horizontal or vertical. These answers
  have been obtained by combining the measured circular and linear
  polarizations in different ways to derive the field
  inclinations. A problem with these combinations is that the circular
  and linear polarizations scale with field strength in profoundly
  different ways. Here, we avoid these problems by using an entirely
  different approach that is based exclusively on the fundamental
symmetry properties of 
the transverse Zeeman effect for observations away from the disk
center without any dependence on the circular
  polarization. Systematic errors are suppressed by the application
of a doubly 
differential technique with the 5247-5250\,\AA\ line pair for
observations with the ZIMPOL-2 imaging polarimeter on the French
THEMIS telescope on Tenerife. For the weakest, 
intranetwork-type magnetic fields, the angular distribution changes
sign with the center-to-limb distance, being preferentially horizontal
limbwards of $\mu$ (cosine of the heliocentric angle) $=0.2$, while
favoring the vertical direction inside this disk position. Since
decreasing $\mu$ corresponds to increasing height of line
formation, this finding implies that the intranetwork fields are more
peaked around the vertical direction in the low to middle
photosphere, while they are more horizontal in the upper
photosphere. The angular distribution is however also found to become
more vertical with increasing flux density. Thus, all facular points
that we have 
observed have a strong preference for the vertical direction for all
disk positions, including those all the way to
the extreme limb. In terms of spatial averages 
weighted by the intrinsic magnetic energy density, these results
are independent of telescope resolution. 
}

\maketitle

\section{Introduction}\label{sec:intro}
Although the relation between the Zeeman-effect polarization and the
magnetic field vector (strength and orientation) has been known for a
century, most explorations of solar magnetic fields have been based on
the longitudinal Zeeman effect alone, which gives us the line-of-sight
component of the field. There have been relatively few attempts to determine the
full field vector \citep[starting with][]{stenflo-stepseverny62} by using the transverse
Zeeman effect as well, but there are very good reasons for
this. Firstly  the measurement noise for weak to moderately strong fields is larger
by  typically a factor of 25 in the transverse field component as
compared with the longitudinal field component, assuming that the
polarization noise is the same for the circular and linear
polarization measurements. In addition to this magnetic
insensitivity, the relation between polarization and field strength is
highly nonlinear (approximately quadratic) in the transverse
Zeeman-effect case, in contrast to the nearly linear field-strength
dependence on the circular polarization. It is this linear response that
allows direct mapping of the line-of-sight component of the
field. What usually goes under the name magnetogram is simply a
map of the circular polarization recorded in the wing of a
Zeeman-sensitive spectral line. 

The nonlinear response of the linear polarization to the magnetic
field creates a serious problem for the interpretation of the
measurements. It is well known from past work with the longitudinal
Zeeman effect that the magnetic structuring continues on scales far
smaller than current resolution limits of solar telescopes
\citep[cf.][]{stenflo-s12aa1}. While the optical average over the 
spatial resolution element leads to a circular polarization that
(because of its linear response) can be interpreted as a magnetic flux, the
meaning of the corresponding optical average for the linear polarization
depends on the poorly known properties of the subresolution magnetic
structuring. It is therefore not so surprising that previous attempts
to combine circular and linear polarization 
measurements to determine the vector magnetic field have led to highly
divergent results. 

Although the circular polarization alone only gives us one component
of the field vector, it is possible to draw conclusions about the
angular distribution of the field directions by examining how the
appearance of the field pattern changes in line-of-sight magnetograms
as we go from disk center to the limb and by comparing the
appearance of magnetograms made in spectral lines formed at different
heights. By comparing magnetograms in photospheric and in
chromospheric lines, \citet{stenflo-giovanelli80} and
\citet{stenflo-jonesgiov83} thus concluded that the largely vertical magnetic field in
facular regions rapidly diverges with height to develop a significant 
horizontal component with the formation of magnetic canopies having base heights of
about 600-800\,km, which are not far above the temperature minimum. 

Since the application four decades ago of the 5250/5247 line-ratio method
for the circular polarization \citep{stenflo-s73}, we know that most of
the flux that we see on the quiet Sun in full-disk magnetograms comes
from strong (kG-type) fields. The main size range for these flux elements is
  estimated to be 10-100\,km \citep{stenflo-s11aa}, which is
  generally below the resolution of current telescopes, but  there have been direct observations of resolved flux tubes since the
  upper tail of the size distribution falls in the domain that can be
  resolved. This was first accomplished by \citet{stenflo-keller92} by
  introducing the technique of speckle polarimetry and more recently
  by \citet{stenflo-lagg10} with the IMaX magnetograph on the
  balloon-borne SUNRISE telescope. Pressure balance with the
  surroundings demands that the  
magnetic field lines of these flux elements must rapidly diverge with
height because of the exponential decrease of the ambient gas pressure and 
therefore develop canopies in the upper layers. At the same time, the 
buoyancy, which also has its origin in the exponential height decrease
of density and pressure, forces the magnetic configuration to stand upright. We
therefore expect the flux from the intermittent flux
elements (most often referred to as flux tubes) to be
preferentially vertical in the lower photospheric layers and that the
angular distribution will widen with height with an increasing
proportion of highly inclined fields. 

Although the kG fields are responsible for a major fraction of the net
magnetic flux that we see in magnetograms with moderately high spatial
resolution, they occupy only a low fraction of the photospheric volume of
the quiet Sun. Through applications of the Hanle effect, we have long
known \citep{stenflo-s82} 
that the remaining fraction is far from empty. On the contrary, it is
seething with an ``ocean'' of flux that can be referred to as
hidden, since it is invisible in medium-resolution
magnetograms because of cancellation 
of the contributions from the opposite magnetic polarities within the
telescope resolution element. However, it starts to emerge in magnetograms
with high resolution, as in the Hinode data \citep{stenflo-bellot12}. Since much
of this field component is likely to be composed of elements that are
expected to be much smaller 
than the atmospheric scale height \citep[with sizes probably less than a few km,
cf.][]{stenflo-s12aa1}, it has been natural to assume that their
angular distribution is nearly isotropic. 

The kG flux tubes and the microturbulent field represent two
complementary idealizations in the description of the underlying
reality. For theoretical reasons, we expect the weakest and strongest
fields to be connected by a continuous field distribution with a
continuous scale spectrum from the nearly global scales to the magnetic
diffusion limit.  In this continuum, we expect the weaker fields to be
more tilted, since they are more easily buffeted by the turbulent or
granular motions. Mixed-polarity fields of intermediate sizes will form
a multitude of small loops, which will be characterized by largely
horizontal fields at their tops. The question is then if all these 
various processes add up such that the total
flux will on average be more vertical or more horizontal. In any case, 
we expect the relative proportion of horizontal
fields to grow as we move up in height
\citep[cf.][]{stenflo-steiner10}. 

In recent years there has been increasing use of
the combination of the transverse and longitudinal Zeeman effects to
derive both the strength and orientation of the magnetic field
vector  with the availability of imaging polarimeters that
can record the full Stokes vector. Computer algorithms for this conversion (often referred to as Stokes
inversion) have been developed and applied not only to active
solar regions, where the S/N ratio of the data is good, but also to
quiet regions, where the linear polarization signal for the
majority of pixels is overwhelmed by noise. 

Recordings of quiet-sun magnetic fields 
with superb spatial resolution  by the SOT/SP instrument on the Hinode
satellite
\citep{stenflo-kosugi07,stenflo-tsuneta08,stenflo-suematsu08} reveal 
that not only vertical but also intermittent patches of horizontal
magnetic fluxes can be found everywhere on the quiet Sun. Initial analysis
of these data using the combination of the Stokes $V$ circular
polarization signals with those of the linear polarization provided by Stokes
$Q$ and $U$ led to claims not only that the angular distribution is
more horizontal than vertical at the photospheric level of formation
of the 6301 and 6302\,\AA\ lines that were used for the disk-center
observations, but also that there is even as much as five times more
horizontal than vertical flux
\citep{stenflo-orozcoetal07,stenflo-litesetal08}. In contrast,
subsequent analysis of the identical Hinode data set led to the
conclusion that the angular distribution is quasi-isotropic
\citep{stenflo-ramos09} or preferentially vertical, although becoming nearly
isotropic in the weak flux-density limit \citep{stenflo-s10aa}. 

In these various investigations, the field inclination angles have
  been derived from different types of combinations of the observed circular and
  linear polarizations. The circumstance in which the obtained results are
  so divergent indicates that there are pitfalls in doing this. In the
  present work, we use an approach that aims at avoiding these
pitfalls. By refraining from any comparison between the linear and
circular polarization amplitudes and by instead basing our analysis
on the linear polarization alone, as observed far away from the center of
the solar disk to break the geometrical symmetry for the
transverse Zeeman effect, we are able to determine for each
center-to-limb position, whether the angular distribution favors the
horizontal or the vertical direction. This method is model independent
in the sense that it only makes use of the fundamental symmetry
properties of the Zeeman effect.

\section{Stokes profile signatures of field
  orientation}\label{sec:signatures}
\subsection{Symmetry properties of the Zeeman effect}\label{sec:filling}
The dependence of the emergent Stokes
$Q$, $U$, and $V$ parameters on field strength $B$ and orientation
(inclination $\gamma$ relative to the line of sight and azimuth
$\chi$ relative to the direction that defines positive $Q$) can be
expressed in the following factorized form\citep{stenflo-book94} as 
\begin{eqnarray}\label{eq:factor} 
Q&= & q_\lambda \,g_Q\,,\nonumber \\ 
U&= & u_\lambda  \,g_U\,,\\ 
V&= & v_\lambda  \,g_V\,,\nonumber 
\end{eqnarray}
where nearly all of the angular dependence is contained in the angular
factors of 
\begin{eqnarray}\label{eq:angscale} 
g_Q&=&\sin^2\!\gamma\,\cos 2\chi\,,\nonumber \\ 
g_U&=&\sin^2\!\gamma\,\sin 2\chi\,,\\ 
g_V&=&\cos\gamma\,, \nonumber 
\end{eqnarray}
while $q_\lambda$, $u_\lambda$, and $v_\lambda$ represent profile
functions that describe the shape of the respective Stokes profile. 
In the optically thin case (weak-line limit), the profile functions
depend on field strength $B$ but not on field orientation, and
$q_\lambda =u_\lambda$.  In the general case when the spectral lines
are optically thick, $q_\lambda$ and $u_\lambda$ can differ 
from each other, but they
are statistically identical for an axially symmetric field
distribution. Each value of $q_\lambda$, $u_\lambda$, and 
$v_\lambda$ also has a weak dependence on field orientation,  
besides field strength, in the optically thick regime. This directional
dependence has its origin both in line saturation
(radiative-transfer effects) and magneto-optical effects and may
mildly modify the shape of the profile functions but not their sign
pattern. 

For absorption lines, the central $\pi$ component of the transverse
Zeeman effect is linearly polarized with the electric vector perpendicular
to the direction of the transverse field component, while the blue-
and red-shifted $\sigma$ component lobes are linearly polarized along
the transverse field. For emission lines, we have the orthogonal 
polarization orientations. The signs of $Q$ and $U$ depend on the direction
that we choose to define positive Stokes $Q$, but regardless of this
definition the $\pi$ and $\sigma$ components of $q_\lambda$ and $u_\lambda$
have opposite signs. 

The relative sign pattern of the transverse Zeeman-effect profiles
(e.g. whether the $\pi$ component is positive or negative) is
determined by the angular functions $g_Q$ and $g_U$. The
weak directional dependence of $q_\lambda$ and $u_\lambda$ does not
significantly 
change this qualitative aspect, although it modifies the shape and
amplitudes. The analysis approach of the
present paper is exclusively based on this qualitative symmetry
property. It therefore does not depend on any assumptions
concerning the field strengths (weak or strong) or line strength
(optically thick line-formation effects). 

Although the present work does not make direct use of the $B$
dependence of $q_\lambda$ and $u_\lambda$, this dependence plays a role in
the understanding of the meaning of the spatial averages of the observed $Q$
and $U$ parameters, since it governs the weighting functions in the ensemble
averages. If the Zeeman splitting is much
smaller than the line width, both $q_\lambda$ and $u_\lambda$ of the
linear polarization are
proportional to $B^2$ or the magnetic energy density, while
$v_\lambda$ of the circular polarization is proportional to $B$ or the
magnetic flux. For the photospheric lines with which we are normally
dealing, 
this weak-field case represents an excellent approximation for
$B\la 0.5$\,kG.  We expect this regime to cover most of the
intranetwork magnetic fields, although our analysis does not depend on
this assumption. 

The concept of using the symmetry properties of Stokes $Q$
distributions was introduced and applied long ago in center-to-limb
observations of Stokes $Q$ 
\citep{stenflo-s87} to constrain the angular distribution of the spatially
unresolved vector magnetic fields together with the field-strength
constraints imposed on these hidden fields by the Hanle 
effect. The implementation of this concept was limited at that time by
being based on observations with 1-pixel detectors 
(photomultipliers). With the availability of high-precision imaging Stokes 
polarimeters it is therefore only now that we are in a position to
exploit the technique more fully.

\subsection{Angular scaling functions in the atmospheric reference frame}\label{sec:refframe}
The angular distribution functions that govern the signs of the
  Stokes $Q$, $U$, and $V$ parameters must be specified in the
atmospheric reference frame relative to the vertical direction in
the atmosphere, while the observed Stokes parameters have been
expressed relative to the line of sight. We therefore need to
convert the angles $\gamma$ and $\chi$ of the Stokes reference frame
to the angle $\theta_B$ of the field vector relative to the
vertical direction and azimuth $\phi$ of the field vector around the
vertical, using the direction that defines
positive $Q$ as projected onto the horizontal plane as the zero point. This conversion
depends on the viewing angle, defined by 
the heliocentric angle $\theta$, which defines the
center-to-limb distance of the observations at the same time. 

With the help of spherical trigonometry, the converted expressions for
the angular scaling factors in Eq.~(\ref{eq:angscale}) become 
\begin{eqnarray}\label{eq:refframe} 
g_Q&=&-\sin^2\!\theta_B\,\sin^2\!\phi\,+\,(\sin\theta\,\cos\theta_B\,-\mu\sin\theta_B\,\cos\phi\,)^2\,,\nonumber \\ 
g_U&=&2\sin\theta_B\,\sin\phi\,\,\,(\sin\theta\,\cos\theta_B\,-\mu\sin\theta_B\,\cos\phi\,)\,,\\ 
g_V&=&\mu\cos\theta_B\,+\,\sin\theta\,\sin\theta_B\,\cos\phi\,, \nonumber 
\end{eqnarray}
where $\mu=\cos\theta$. The symmetry direction that defines positive
Stokes $Q$ and the zero point of the azimuths is the radius vector
from disk center to the point of observation. 

Next, we follow \citet{stenflo-s87} and characterize the distribution
over all the field inclination angles $\theta_B$ by the function, 
\begin{equation}\label{eq:a}
f_a\,\sim\mu_B^a\,,
\end{equation}
which is defined by the single free parameter $a$. Here, 
$\mu_B=\cos\theta_B$. The azimuth distribution is assumed to be
axially symmetric (independent of $\phi$). The normalization constant is fixed by the
requirement that the integration over all directions should give
unity. 

We can now conveniently classify the nature of the various angular
distributions in terms of parameter $a$: An isotropic distribution is
described by $a=0$, while a flat ``pancake'' distribution consisting
of fields that are confined to the horizontal plane is 
represented by $a=-1$. As $a$ increases, the distribution 
increasingly peaks around the vertical direction. In the limit of
infinite $a$, all fields are exactly vertical. 

Let us already stress here that the results of the present paper, which are based on the
qualitative symmetry properties of the $g_Q$ distributions, do not 
depend on the particular parametrization described by
Eq.~(\ref{eq:a}), although the particular choice of the 
$\mu_B^a$ distribution is made for mathematical and conceptual
convenience. 

\begin{figure}
\resizebox{\hsize}{!}{\includegraphics{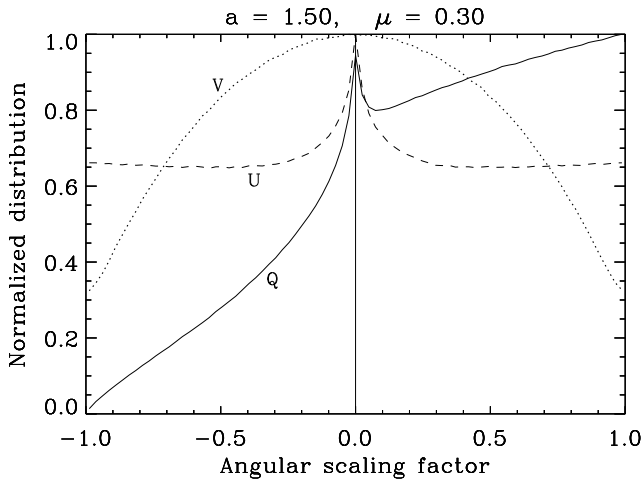}}
\caption{Distribution of the angular scaling factors $g_{Q,U,V}$ of
  Eq.~(\ref{eq:refframe}),  representing the geometric scalings of
  Stokes $Q$ (solid line), $U$ (dashed  line), and $V$ (dotted
  line). The scalings were computed by Monte Carlo simulations for the
  choice of an angular 
  distribution parameter $a=1.5$ and a viewing angle defined by
  $\mu=0.3$. 
}\label{fig:anghisto1}
\end{figure}

We have used Monte Carlo simulations to explore the properties of the
$g_{Q,U,V}$ distributions that result for a given choice of $a$
(angular-distribution parameter) and viewing angle (defined by
$\mu$). Thus, each of the values for $g_{Q,U,V}$ is based on a pair of
independent Monte Carlo samplings: one for $\phi$ based on a flat
distribution and one for $\theta_B$ based on the distribution of
Eq.~(\ref{eq:a}). We use 20 million of these samples for each ($a$, $\mu$)
combination to obtain smooth distribution functions. Figure
\ref{fig:anghisto1} shows an example for the choice of $a=1.5$
(representing a distribution that favors vertical fields) and $\mu=0.3$
(representing a typical $\mu$ value of our observational data set). 

The fundamental symmetry property is clearly brought out by
Fig.~\ref{fig:anghisto1}: While the $g_U$ and $g_V$ distributions are
symmetric relative to positive and negative values, the $g_Q$ distribution is
not, as it strongly favors positive values. Therefore, an average over
the distributions of $Q$, $U$, or $V$ values would give zero for $U$ and
$V$ (assuming that the statistical sample used is sufficiently large),
while the average of $Q$ would be nonzero. This symmetry property is
valid for all viewing angles with the singular exception of disk
center ($\mu=1.0$), where the $Q$ average also becomes zero because of  
symmetry around the vertical direction. 

\begin{figure}
\resizebox{\hsize}{!}{\includegraphics{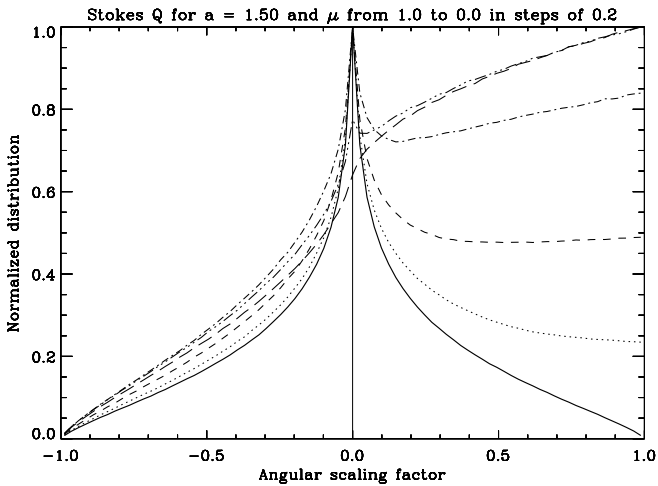}}
\caption{Distribution of the angular scaling factor $g_Q$ for
  $a=1.5$ and a sequence of $\mu$ values, decreasing from the
  disk-center value 1.0 (the symmetric solid line) to the extreme limb
  value of 0.0 in steps of 0.2. As $\mu$ decreases, the curves in the
  right-hand part of the diagram become increasingly elevated. 
}\label{fig:anghisto2}
\end{figure}

The asymmetry of the $g_Q$ distribution increases monotonically from
zero as we go from disk center to the limb, as illustrated in
Fig.~\ref{fig:anghisto2}, for the same choice of angular distribution
parameter as in
Fig.~\ref{fig:anghisto1} ($a=1.5$). For negative values of
$a$, the asymmetry goes in the opposite direction in favor of negative
values. The more $a$ deviates from zero, the more pronounced the
asymmetry becomes. 

As we average $g_{Q,U,V}$ in Eq.~(\ref{eq:refframe}) 
over the angular
distribution defined by Eq.~(\ref{eq:a}), we get zero for $g_{U,V}$,
while the average $\bar g_Q$ of $g_Q$ is 
\begin{equation}\label{eq:gqave}
\bar g_Q\,=\,{a\over a+3}\,(1-\mu^2)
\end{equation}
\citep{stenflo-s87}. Note that $1-\mu^2=(r/r_\odot)^2$, where $r/r_\odot$ is the radius
vector in units of the disk radius $r_\odot$. Eq.~(\ref{eq:gqave})
directly shows how the sign of $\bar g_Q$ determines whether the
distribution is more vertical or more horizontal.

\subsection{Relation between the $g_Q$ and $Q$ 
  distributions}\label{sec:qobs}
While the observed Stokes $Q$ scales with both $q_\lambda$ and $g_Q$,
the profile function $q_\lambda$ has a fixed relative sign pattern
between its $\pi$ and $\sigma$ components, although its detailed shape
and amplitude scale with the field strength (in the weak-field regime
in proportion to $B^2$). This sign pattern is either of the form $-+-$ or
$+-+$. These two forms are governed by the
angular function $g_Q$.  

In our analysis approach, we deal with ensemble averages of field
distributions. These distributions are functions of both field
strength $B$ and field orientation. If the $B$ distribution were
stochastically independent of the angular distribution, then the sign
of the average ${\bar  Q}$ of the observed $Q$ 
distribution would be determined exclusively by the sign of
$\bar g_Q$, and spatial averaging over $Q$ would unambiguously
determine whether the fields are more horizontal or more
vertical. However, the angular distribution is $B$ dependent  as will
be shown in the present analysis, in the sense that the distribution 
becomes more peaked around the vertical direction as the flux density
increases, as found earlier from 
Hinode quiet-sun disk-center analysis \citep{stenflo-s10aa}. We know that
there is a wide range of field strengths that coexist on the
quiet Sun from near-zero to kG fields. In terms of our $a$ parameter
that characterizes the angular distributions, this implies that there
is a coexistence of a range of $a$-values. 

To understand the meaning of the observed ${\bar Q}$
averages, let us subdivide the field-strength distribution in
field-strength bins and consider the total vector field distribution
(of both field strength and orientation) as a linear
superposition of angular distributions with a different value of $a$
for each field-strength bin. The average ${\bar Q}$ is then
obtained by first forming the angular averages of $Q$ for
each field-strength bin and then the weighted sum of these
averages over all the field-strength bins with $q_\lambda$ as
weights is computed. Since $q_\lambda\sim
B^2$ is a good approximation for field strengths below about 0.5\,kG,
the weighting function is approximately proportional to the
pixel-average of the magnetic energy density. However, our analysis
does not depend on any assumption concerning the validity of this
approximation. 

The meaning of the sign of ${\bar Q}$ depends on whether we
refer $Q$ to the $\pi$ or $\sigma$ component amplitude of the
transverse Zeeman effect (see below). For the present
discussion, let us assume that positive ${\bar Q}$ means that 
the field distribution is preferentially vertical under the
assumption of a $B$-independent angular parameter $a$. In the
generalized framework of a $B$-dependent $a$, a positive ${\bar Q}$
then means that the distribution when weighted (approximately in terms
of the magnetic 
energy densities) is preferentially vertical, while it is
preferentially horizontal when ${\bar Q}$ is negative. 

Since this weighting favors the contributions from the higher flux
densities like facular points and, as we will see below, these fluxes
are always strongly peaked around the vertical direction for all disk
positions (all $\mu$ values), the contribution of the weak
fluxes may easily get overwhelmed and masked by the domination of
concentrated fields. As the main focus of the present work is
to explore the angular distribution properties of the weakest fluxes,
since they represent the only viable candidates for horizontal fields,
we have particularly selected recordings in the most quiet, non-facular
solar regions, where all the fluxes covered by the spectrograph slit
are extremely weak (with very few pixels having $Q/I$ amplitudes in
excess of 0.2\,\%\ and $V/I$ in excess of 0.5\,\%). The measurement of
these weak polarization signatures has become possible thanks to the
ZIMPOL technology, with which the polarimetric precision is almost
exclusively limited by photon statistics. The 1-$\sigma$ error in
the determination of the $Q/I$ amplitude is on the order of 0.01\,\%\ for
each spatial pixel and is reduced by an order of magnitude when
averaging over all pixels along the slit. By isolating these weak
fluxes, the observationally
determined averages ${\bar Q}$ will avoid the dominance of
the higher flux densities and more closely approximate the angular-distribution
properties in the weak flux limit, which is representative of what we may call
the intranetwork fields. 

\subsection{Influence of the telescope resolution}\label{sec:telres}
The averaging over the values of $Q$ that gives us ${\bar Q}$ for a given area
  of the solar disk can be done in two different ways:
(1) Optical integration over the area, or (2) numerical integration over the
different resolved spatial pixels in the selected area. In practice, 
there is always a combination of the two, since the quiet-sun
magnetic fields are at best only 
partially resolved. As the integration over Stokes $Q$ represents a linear
superposition of the contributions from each infinitesimally small
solar surface element, the results of an optical and a numerical
integration however are identical. This would not at all have been the case if
we would have averaged over flux densities or field
strengths instead, since the relation between Stokes $Q$ and field strength is
very nonlinear. As the observed $Q$
is however proportional to the number of polarized photons,
optical and numerical summation are equivalent. 

The consequence of this equivalence is that the average ${\bar Q}$
over a given area of the Sun's surface is independent of 
telescope resolution. The average would not be 
different if we would use the substantially higher spatial resolution
of Hinode, or any hypothetical future telescope with far higher
angular resolution. 

Let us for clarity stress that the resolution independence that
  we talk about here exclusively refers to the {\it ensemble
    averages}. The magnetic structuring that is directly revealed by
  the observations is, of course, extremely dependent on telescope
  resolution. With each new generation of solar telescopes, the solar
  images reveal a new world of structures that was not visible before. In the
  Stokes images presented in the present paper and elsewhere,
  structures of varying amplitudes, signs, and sizes are seen
  everywhere, and their visibility depends on telescope
  resolution. This dependence does not in any way contradict our
  arguments concerning independence of telescope resolution for the
  ensemble averages. The only assumption that is involved here is that
the statistical sample used when calculating these ensemble averages
is sufficiently large and representative. This is not a principle
limitation, since it can be improved through increase of the size of
the data base.

Similar resolution independence has been encountered 40 years ago with
the Stokes $V$ 5250/5247 line ratio technique (which revealed the
intermittent kG nature of quiet-sun magnetic flux) and 30 years ago
with the application of the Hanle depolarization technique (which
revealed the existence of the hidden, turbulent magnetic
field). Since the
unique power of all these methods is their independence of 
telescope resolution, observational applications of them give priority to 
polarimetric precision combined with high spectral resolution, while
high spatial resolution is secondary.

\section{Observations and reduction technique}\label{sec:obs}

\subsection{ZIMPOL-2 at THEMIS}\label{sec:zimpol2}
The data set used here was recorded on Tenerife with the ZIMPOL-2 polarimeter of
ETH Zurich that was installed at the French THEMIS telescope, which
has a 90\,cm
aperture and is nearly polarization-free at the location in the
optical train where the polarization is being analyzed. Two observing
campaigns with ZIMPOL on THEMIS have been carried out so far: one in
2007 (July 30 - August 13, 2007) and the other in 2008 (May 29 - June 12,
2008). The main aim was to explore the spatial structuring of the
Second Solar Spectrum with its Hanle effect in different prominent
spectral lines, but some recordings of the Zeeman effect were also
carried out, in particular for the 5247-5250\,\AA\ data set used here. 

The ZIMPOL (Zurich Imaging Polarimeter) technology
\citep{stenflo-povel95,stenflo-povel01,stenflo-gandetal04} solves the
compatibility problem between fast (kHz) polarization modulation and
the slow read-out of large-scale array detectors by creating fast
hidden buffer storage areas in the CCD sensor. This allows the
photocharges to be cycled between four image
planes in synchrony with the fast modulation. Linear combinations of
the four image planes give us the simultaneous images of the four
Stokes parameters. The images of the fractional polarizations, $Q/I$,
$U/I$, and $V/I$, are free from both gaintable and seeing noise, because
the flat-field effects divide out when the fractional
polarizations are formed (since the identical pixels are used for the four image
planes), and because the modulation is substantially faster than the
seeing fluctuations. 

The ZIMPOL-2 version creates the four fast buffers by placing a mask
on the CCD (in the manufacturing process), such that three pixel rows
are covered for buffer storage for every group of four pixel rows, while
one is exposed. To prevent loss of 75\,\%\ of the photons that
would fall on the masked part of the CCD, an array of cylindrical
microlenses is mounted on top of the pixel array. Each cylindrical
lens has the width of four pixels and focuses all the light on the
unmasked pixels, thereby eliminating the light loss. 

To modulate all four Stokes parameters simultaneously at about 1\,kHz, we use two
phase-locked ferro-electric liquid crystal (FLC) modulators in
combination with a fixed retarder, followed by a fixed linear
polarizer. By optimizing the relative azimuth angles of the three
components in front of the polarizer, the modulator can be made
sufficiently achromatic to be able to simultaneously modulate the
Stokes parameters with high efficiency in two widely separated
wavelength windows. Since our campaign employed two ZIMPOL-2 CCD
sensors to exploit the opportunity offered at THEMIS to
simultaneously observe in different parts of the spectrum, we needed
this semi-achromatic optimization. The optimized settings led to a 
Mueller matrix of the modulator package that is highly nondiagonal in
different ways at the selected wavelengths, but this is of no concern,
since the matrix is always fully determined by the polarization
calibration procedure and
then numerically diagonalized. The only criterion is to simultaneously
optimize the modulation efficiency in the four Stokes parameters. 

The field of view for the observations was approximately 70\,arcsec
(along the slit) in the spatial direction and about 5.1\,\AA\ in the
wavelength direction. The array size was 770 pixels in the wavelength
direction multiplied by $4\times 140 =560$ pixels in the spatial
direction. Since the grouping of the pixel rows by the mask and
microlenses is done in the slit direction, the effective number of
spatial resolution elements was 140. The size of this effective
resolution element corresponded to 0.5\,arcsec on the Sun, which means
that the effective (seeing-free) spatial resolution was limited to
1\,arcsec. Seeing compensation was done with an active-optics tip-tilt
system. The pixel size in the wavelength direction was
6.6\,m\AA. The slit width was about 1.0\,arcsec in the solar image,
which corresponds to about 40\,m\AA\ in the spectral focus. 

The spectrograph slit was always oriented perpendicular to the radius
vector, i.e., parallel to the nearest solar limb. In all our
reductions and figures, the orientation of the Stokes system has been
defined such that the positive $Q$ direction is parallel to the
slit. In the 2-D spectral images, positive polarization signatures 
are brighter, and negative ones are darker than the surroundings. 

Since the polarimetric sensitivity of the ZIMPOL system is practically
only limited by the Poisson statistics of the collected
photoelectrons, the S/N ratio scales with the square root of the
effective integration time. Since polarimetric precision had the
overriding priority over spatial and temporal resolution for the
present type of work, very long effective integration times were used,
usually 20-30\,min, which is typical of explorations of the Second Solar
Spectrum and the Hanle effect. These long integrations are also
much needed when dealing with the miniscule signatures of the
transverse Zeeman effect on the quiet Sun.

\subsection{5247-5250 data set}\label{sec:dataset}
The optimum region in the visible spectrum in terms of both sensitivity to
the Zeeman effect and highest degree of model
independence in magnetic-field diagnostics is the 5247-5250\,\AA\
range that includes the line pair Fe\,{\sc i} 5247.06 and
5250.22\,\AA. This line pair is unique in several respects: As the two
lines belong to the same multiplet (no.~1 of neutral iron), have the same line
strength, and have the same excitation potential, they are formed in the same way in
the solar atmosphere, in contrast to most other line combinations 
like the Fe\,{\sc i} 6301.5 and 6302.5\,\AA\ lines, which differ
significantly in their thermodynamic properties and heights of
formation. The 5250\,\AA\ line has a Land\'e\ factor of 3.0,
unsurpassed by any other relevant line in the Sun's spectrum. It is
therefore one of the most Zeeman-sensitive lines in the Sun's visible
spectrum. The only significant difference between the two lines of the
pair is their Land\'e\ factors: For the  5247.06\,\AA\ line the
effective Land\'e\ factor is 2.0, which is still large enough to also make this line
highly Zeeman sensitive but is significantly different from 3.0, which
makes this line combination an ideal ``magnetic line ratio'' for differential,
nearly model-independent diagnostics. 

The discovery
of the kG nature of quiet-Sun magnetic flux was made by 
determining the differential Zeeman saturation from the ratio of the
circular polarizations in these two lines \citep{stenflo-s73}. Their
identical line-formation properties allows a clean separation of the
magnetic-field effects from the thermodynamic effects, although the
sizes of the responsible flux elements are far lower than the
telescope resolution. This insight laid the foundation for the
construction of semi-empirical flux tube models, which were
largely based on multiline analysis of recordings with an FTS
(Fourier Transform Spectrometer) polarimeter \citep[cf.][]{stenflo-sol93}. 

Here, we will not use the 5247-5250 line pair in the traditional
way by forming Stokes $V$ or Stokes $Q$ line ratios, but we instead 
determine differential effects between the two lines when computing
the spatial averages of Stokes $Q$. To eliminate the possible
infiltration of subtle but systematic polarization effects at the level below
0.01\,\%, e.g. weak polarized fringes, we determine not only the
$Q$ amplitude difference between the $\pi$ and $\sigma$ components of
the transverse Zeeman pattern but also the difference between
the two lines of these quantities. It is the use of these doubly
differential quantities that allows an extraordinary precision in the
determination of the angular distributions. 

The recordings of the 5247-5250\,\AA\ range with ZIMPOL-2 at
THEMIS were carried out on June 9-10, 2008. On the first day, a
sequence of quiet-sun recordings at $\mu$ positions 0.1, 0.2, 0.3,
0.4, and 0.5 from the heliographic N pole along the central meridian
were performed to provide us with a center-to-limb sequence of the most
quiet Sun. Inspection of the recordings showed that none of the slit
positions contained contributions from any facular points. All of them
represented what we would call intranetwork. On the second day, 
recordings in a number of limb regions with various types of facular
points (from tiny to medium size) and small spots and additional
recordings in other quiet regions were made. A total of 14 
high-quality Stokes spectra of the 
5247-5250\,\AA\ range were thus secured. 

\subsection{Evidence of vertical or horizontal fields through visual inspection}\label{sec:evidence}

\begin{figure}
\resizebox{\hsize}{!}{\includegraphics{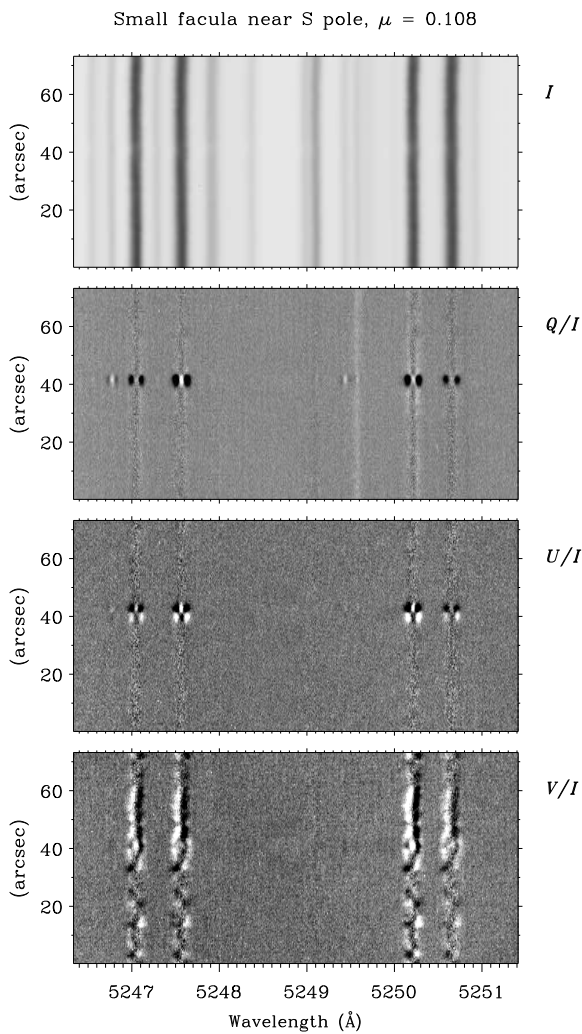}}
\caption{Stokes spectral images obtained with the slit crossing a small facula at
$\mu=0.108$ near the heliographic S pole. The recording was made with
ZIMPOL-2 at THEMIS on June 10, 2008. The transverse Zeeman-effect
signatures in $Q/I$ show that the facular magnetic field is nearly vertical.
}\label{fig:facula}
\end{figure}

Figure \ref{fig:facula} shows one example of such a recording (from
June 10, 2008) across a small facula close to the solar limb at
$\mu=0.108$ near the heliographic S pole. The facula stands out in the
$Q/I$ and $U/I$ images with its spatially localized intense and symmetric
polarization signatures from the transverse Zeeman effect, not only in
the 5247.06 and 5250.22\,\AA\ lines, but also in the Cr\,{\sc i}
5247.57 and Fe\,{\sc i} 5250.65\,\AA\ lines with Land\'e\ factors of
2.5 and 1.5, respectively. In contrast, the antisymmetric $V/I$
profiles from the longitudinal Zeeman effect are not that localized
but are conspicuous along most of the slit. Much of the reason for
this difference in appearance between the transverse and longitudinal
Zeeman effect has to do with their profoundly different relations
between field strength and polarization. While this relation is nearly
linear in the case of the circular polarization, it is highly
nonlinear and approximately quadratic for the linear
polarization. This leads to a relative magnification of the
linear polarization in strong-field and relative suppression
in weak-field regions. 

When the sign of the polarization signatures of the transverse Zeeman
effect with the central $\pi$ component of $Q/I$ is positive while
the $\sigma$ components in the line wings are negative, this implies that the
transverse component of the facular magnetic field is 
aligned with the radius vector, which for limb observations is
evidence 
that the field is nearly vertical. If the transverse field had been
perpendicular to the radius vector, the signs would have been
reversed. For transverse fields that are $\pm 45^\circ$ to the radius
vector, there would be no signal in $Q/I$, only in $U/I$ instead. The $U/I$
image shows a sign change across the facula. The signs are 
balanced, so that the spatially averaged $U/I$ almost vanishes. The
presence of nonzero $U/I$ signatures shows that the facular field is
not strictly vertical but has a significant angular spread around the
vertical direction. 

This sign behavior has been found for all faculae and spots for all
center-to-limb positions that we
have observed. Mere inspection of
the Stokes spectra thus allows us to
conclude that all these magnetic features have 
magnetic field distributions that peak around the vertical
direction. Since the height of line formation increases as we move
towards the limb, the field must retain the strong
preference for a vertical orientation over the entire height range
covered by our various $\mu$ values. 

When we carefully inspect the slit region outside the small
facula in Fig.~\ref{fig:facula} in the most Zeeman-sensitive line
(5250.22\,\AA), we notice however a faint polarization signature that has a
sign opposite to that of the facula almost all along the slit with
positive $\sigma$ components and negative $\pi$ components. This is the 
signature for predominantly horizontal fields. It indicates that there
is a profound dependence of the angular distribution on flux density: While
the stronger isolated flux concentrations, like faculae, are found to be nearly
vertical at all disk positions, the weakest background fields on the
quiet Sun are preferentially horizontal near the solar limb. 

\begin{figure}
\resizebox{\hsize}{!}{\includegraphics{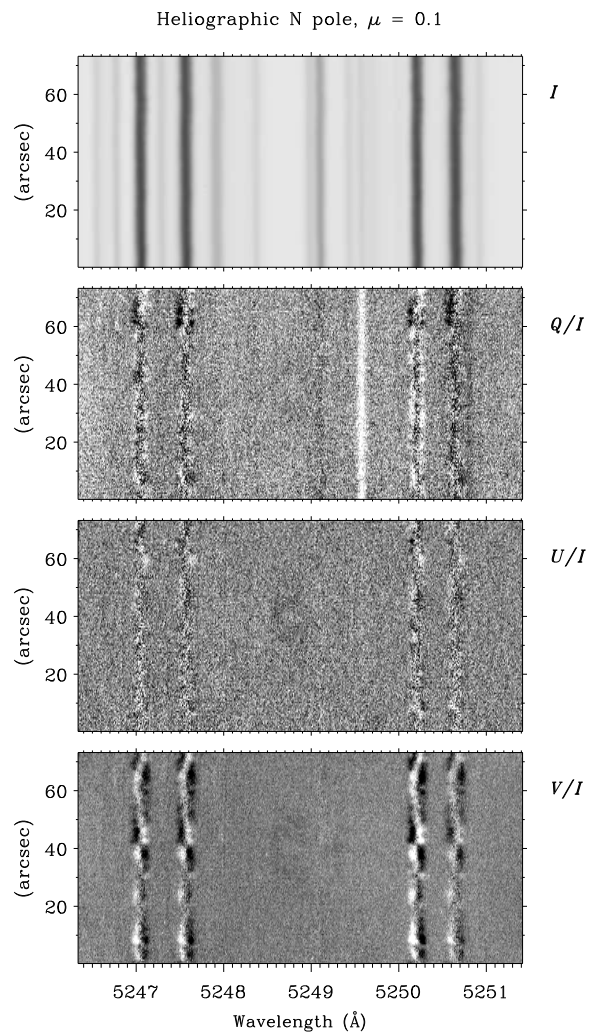}}
\caption{Stokes spectral images of a very quiet region (devoid of faculae) at
  $\mu=0.1$ near the heliographic N pole recorded with ZIMPOL-2 at
  THEMIS on June 9, 2008. Direct inspection reveals that the weak
  flux densities are preferentially horizontally oriented and that
  the great majority of the corresponding magnetic elements are
  unresolved. 
}\label{fig:mu01}
\end{figure}

We therefore need to turn our full attention to the weak background
fields by selecting recordings of the most quiet regions, which are devoid of
any facular points. Figure \ref{fig:mu01} shows a recording made
at $\mu=0.1$ near the heliographic N pole. In contrast to
Fig.~\ref{fig:facula} we can set the grey-scale cuts much lower to bring out the weakest
signals, since  there are no strong polarization features. Inspection 
of $Q/I$ for the 5250.22\,\AA\ line confirms the 
impression from Fig.~\ref{fig:facula} that the $\sigma$ components are
predominantly positive and the $\pi$ component negative, as expected for
preferentially horizontal fields. 

Readers unfamiliar with the Second Solar Spectrum (the linearly
polarized spectrum that is exclusively formed by coherent scattering
processes) may be surprised by the bright emission-like line to the
left of the 5250\,\AA\ line in the $Q/I$ panel. It was also seen in
Fig.~\ref{fig:facula} but stands out in much higher contrast in
Fig.~\ref{fig:mu01} because of the lower setting of the grey-scale
cuts. Like most rare earth lines in the Sun's spectrum the Nd\,{\sc
  ii} 5249.58\,\AA\ line (of singly ionized neodynium) exhibits strong
scattering polarization (with the electric vector oriented parallel to
the solar limb)  \citep[cf.][]{stenflo-sk97} at 
the same time as it shows no trace of the Zeeman effect --- an enigmatic
behavior that is not yet understood. Near the limb, the continuous
spectrum is also significantly polarized, mainly by coherent
scattering in the distant wings of the Lyman series lines and by
Thomson scattering \citep{stenflo-s05}. The transverse Zeeman-effect
signatures therefore sit on an elevated $Q/I$ continuum, which must be
accounted for in the quantitative analysis. 

\begin{figure}
\resizebox{\hsize}{!}{\includegraphics{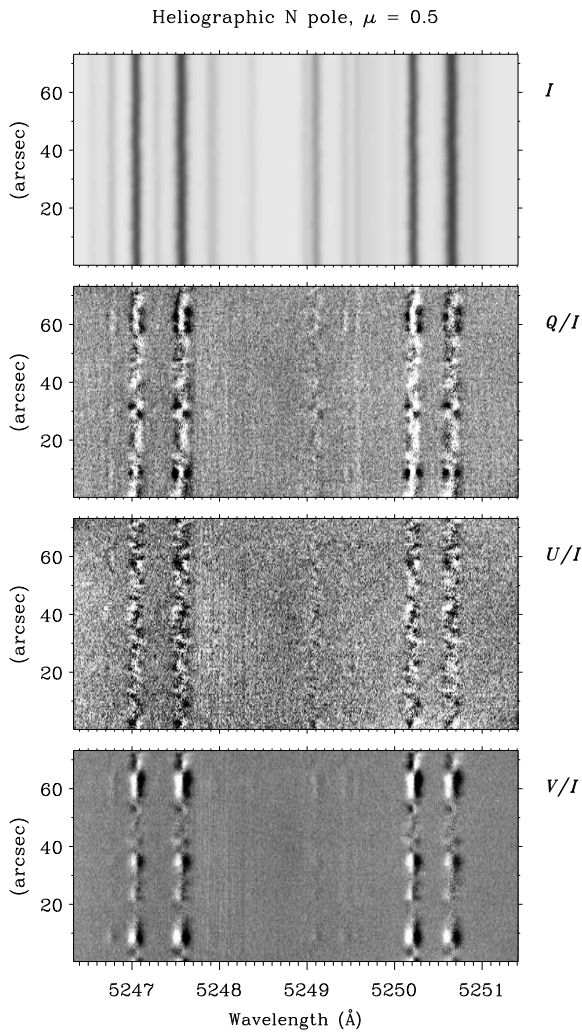}}
\caption{Same as for Fig.~\ref{fig:mu01}, except that the recording
  was made at a $\mu$ position of 0.5 near the heliographic N
  pole. Visual inspection reveals that for this larger $\mu$ value
  the weak flux densities have changed to be more
  vertically oriented. 
}\label{fig:mu05}
\end{figure}

As we move away from the limb, the angular distribution of the
background fields becomes preferentially vertical, as shown by
Fig.~\ref{fig:mu05} for a very quiet region at $\mu=0.5$ near the
heliographic N pole. Visual inspection of $Q/I$ of the
5250.22\,\AA\ line reveals that it is now the $\pi$ component that is
predominantly positive, while the $\sigma$ components are more
negative, which is the signature for predominantly vertical fields. 

Direct inspection of Figs.~\ref{fig:mu01} and \ref{fig:mu05} also allows us to 
conclude that the magnetic elements that collectively
constitute the background field are much smaller than the telescope
resolution. One telltale signature of the presence of multiple distinct magnetic
elements within each spatial resolution element is the highly
anomalous $U/I$ profile shapes 
in Fig.~\ref{fig:mu05} with no resemblance of the
well-known symmetry properties of the transverse Zeeman effect
(approximate symmetry around line center). This can easily be
understood in terms of the superposition of spatially unresolved
magnetic elements, of which  each has slightly different Doppler
shifts because of their different positions within the solar
granulation. It is well known from explorations of Stokes $V$ profiles
\citep[e.g.][]{stenflo-sig01} 
that the superposition of the contributions from unresolved magnetic
elements with different relative Doppler shifts is the main cause of 
the observed anomalous Stokes $V$ profile shapes. This
  superposition does not have to be for different unresolved
  elements in the transversal plane. Anomalous $V$
  profiles are also produced by 
  correlated velocity and magnetic field gradients {\it along} the line of
  sight \citep{stenflo-illing75,stenflo-auer78}. Anomalous $Q$ and $U$
profiles are produced in the same way.

\subsection{Extraction of Stokes parameters}\label{sec:extract}
After having demonstrated how far-reaching and model-independent conclusions
can be made from pure visual inspection of the Stokes images, we now
turn to the quantitative analysis of the images. Since we are dealing
with very weak polarization signals, most of which have amplitudes in
the range of 
0.01 - 0.1\,\%, and the 1-$\sigma$ noise level is on the order of 0.01\,\%\
per spatial pixel, it is imperative to use a technique
for the extraction of the Stokes profile amplitudes, which does not
lead to skewed distributions for amplitudes comparable to or smaller
than the noise, but which have a symmetric Gaussian error
distribution. This ensures that the observational histogram (PDF) of Stokes
profile amplitudes represents the intrinsic, noise-free histogram
convolved with the Gaussian noise distribution. 

This technique was developed and implemented in the analysis of
quiet-sun Hinode SOT/SP spectra by \citet{stenflo-s10aa}, and it will
be applied here as well. An average Stokes
spectrum that has a high S/N ratio is used to construct templates. The blue
and red $\sigma$ component lobes and the $\pi$ component lobe are cut
out and amplitude normalized to become the templates for iterative least
squares fitting of the Stokes spectrum for each spatial pixel. Before
fitting, the spectrum is Doppler shifted and interpolated to the
wavelength scale used for the template spectrum. The only free
parameter in the least squares fit is the amplitude of the respective
lobe, which is determined together with its standard error. This fitting
procedure is extremely robust with immediate and entirely unique
convergence. The error distribution is symmetric and Gaussian. 

\begin{figure}
\resizebox{\hsize}{!}{\includegraphics{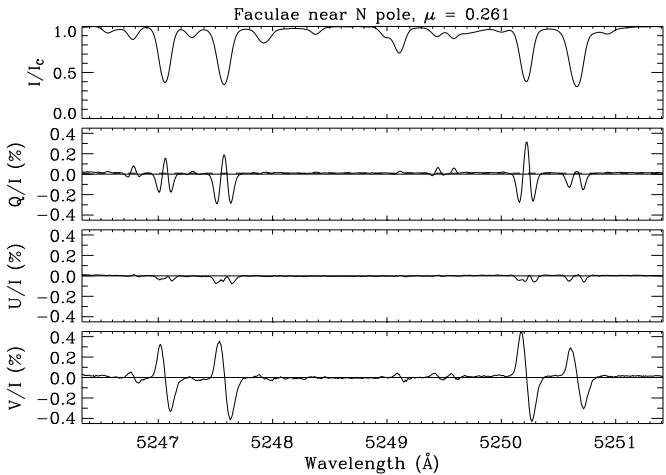}}
\caption{Slit average of the Stokes profiles recorded on June 10,
  2008 in a facular region at $\mu=0.261$ near the heliographic N
  pole. Note the signature of a vertical field: positive $\pi$
  components in $Q/I$, almost zero signal in the spatial average of
  $U/I$. 
}\label{fig:maskspec}
\end{figure}

For the template spectrum we use the average along the spectrograph
slit for a recording across faculae at $\mu=0.261$ near the
heliographic N pole. This averaged spectrum is shown in
Fig.~\ref{fig:maskspec}. Since $U/I$ changes sign along the slit with
a balanced sign distribution, the slit average of $U/I$ is
nearly zero.  In contrast $Q/I$ consistently has the same sign
along the slit, since the fields remain nearly vertical.   

Since the $Q/I$ and $U/I$ relative profile shapes from the
transverse Zeeman effect are identical on average, the lobe templates that
we cut out from the $Q/I$ profiles in Fig.~\ref{fig:maskspec} also
serve as templates for least-squares fitting of the $U/I$ lobes. While
the $V/I$ profiles have two lobes (the blue and red $\sigma$
components), the $Q/I$ and $U/I$ profile have three lobes to fit
(because of the additional $\pi$ component). 

Let us stress here that we make no assumption that the actual
  Stokes profiles resemble the template profiles in
  Fig.~\ref{fig:maskspec}. As the amplitudes of each of the three
  profile lobes of $Q$ and $U$ (the $\pi$ and the two $\sigma$
  components) are determined independently of each other, the profiles
are allowed to have any anomalous balance between these components,
like single-lobe profiles, same-sign lobes, etc. The determination is
independent of the S/N ratio of the data, since each lobe amplitude is
determined with an unbiased, symmetric, and well-defined Gaussian
error distribution. This property represents the special strength of
our extraction technique.

\section{Angular distributions from the CLV of Stokes $Q$}\label{sec:clv}

\subsection{Conclusions from direct visual inspection}\label{sec:inspect}

\begin{figure}
\resizebox{\hsize}{!}{\includegraphics{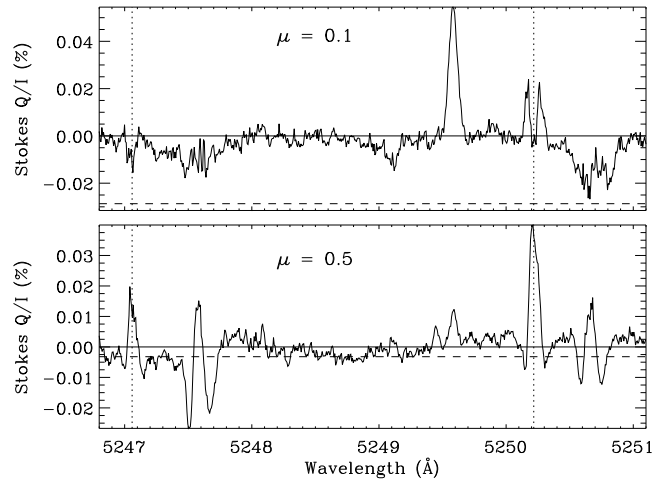}}
\caption{Slit averages of the 2-D $Q/I$ spectra in Figs.~\ref{fig:mu01} and
\ref{fig:mu05}. The zero point of the polarization scale, which should
lie at the level of the dashed line, has been
shifted to represent the continuum level, since the amplitudes of the
Zeeman effect are measured relative to the continuum. The dotted vertical
lines mark the line centers of the 5247.06 and 5250.22\,\AA\
lines. The reversal of the sign of the transverse Zeeman effect from
$\mu=0.1$ to $\mu=0.5$ shows that the angular distribution changes
from predominantly horizontal to vertical as we move away from the
limb. 
}\label{fig:slitave}
\end{figure}

In Sect.~\ref{sec:evidence}, we concluded from visual inspection of
Figs.~\ref{fig:mu01} and 
\ref{fig:mu05} that the $Q/I$ profiles had opposite signs  on average 
in the two figures and that therefore the weak background magnetic
fields on the quiet Sun change from being predominantly horizontal
near the extreme limb to become preferentially vertical when we move
away from the limb. This implies a corresponding height variation in 
the angular distribution, a property that is brought out more clearly in
Fig.~\ref{fig:slitave}, which shows the spectra of Figs.~\ref{fig:mu01} and
\ref{fig:mu05} after averaging along the spectrograph slit. Since the
linear polarization of the transverse Zeeman effect sits on top of an
elevated continuum that is polarized by processes (Rayleigh and Thomson
scattering) that are irrelevant to our Zeeman-effect analysis, we have
shifted the zero point of the $Q/I$ 
polarization scale so that it represents the continuum, relative to
which the amplitudes of the line-profile lobes that are caused by the Zeeman
effect are measured. The true zero point of the polarization scale
is indicated by the dashed line in the two panels. Since the
scattering polarization increases steeply towards the limb, the shift
is much larger for $\mu=0.1$ than for $\mu=0.5$. Similarly, the
polarization amplitude of the neodynium line at 5249.58\,\AA\
decreases steeply as we move away from the limb. 

Figure \ref{fig:slitave} clearly shows that  the $\pi$ component for $\mu=0.1$ points
in the negative direction and the $\sigma$ components in the positive
direction, which indicates a preference for horizontal fields. For
$\mu=0.5$ the signs  however are reversed and represent 
preferentially vertical fields. 

Careful inspection of Fig.~\ref{fig:slitave} indicates that the
continuum level that serves as our reference level for the Zeeman
effect is not entirely flat but there seem to be large-scale
fluctuations on the order of 0.005\,\%, possibly beacuse of weak polarized
fringes. To eliminate errors that are caused by this effect we will not describe
the transverse Zeeman effect in terms of the amplitudes of the $\pi$
and $\sigma$ components alone, but instead in terms of the {\it
  amplitude difference} between 
the $\pi$ component and the average of the blue and red lobe $\sigma$
components. This difference is insensitive to the level of the
continuum or zero point of the polarization scale. By using the
average of the blue and red lobe $\sigma$ components, we effectively
symmetrize the $Q/I$ profiles to avoid issues with the anomalous
profiles that are caused by the superposition of subpixel structures with
different Doppler shifts. The use of these differential techniques is
crucial for all diagnostic work to isolate the
phenomenon of interest from possible contamination from a multitude of
undesired effects.

\subsection{Histogram properties}\label{sec:histo}

\begin{figure}
\resizebox{\hsize}{!}{\includegraphics{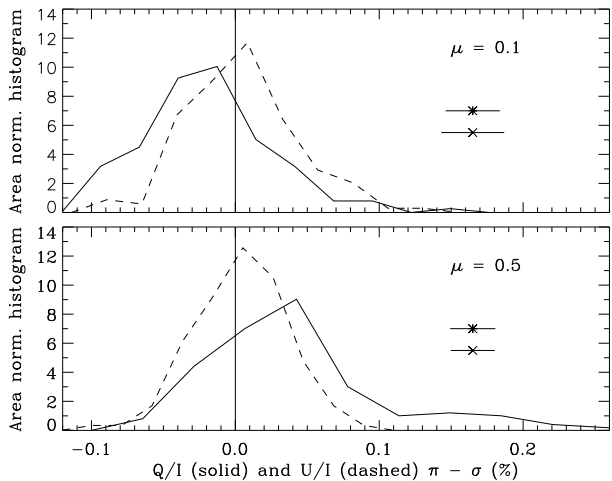}}
\caption{Histograms of $Q/I$
  (solid) and $U/I$ (dashed) for the 5250.22\,\AA\ line, which are sampled along
  the spectrograph slit for 
  the two Stokes spectra illustrated in 
  Figs.~\ref{fig:mu01} and \ref{fig:mu05}. The measured quantity is
  the {\it difference} between the $\pi$ component
and the average of the two $\sigma$ components. The error bars for a
single pixel are
indicated (with asterisk representing $Q/I$, cross $U/I$). While the
$U/I$ distributions are symmetric around zero, the $Q/I$ distributions
are skewed in favor of horizontal fields in the upper panel and in favor
of vertical fields in the bottom panel. 
}\label{fig:pshisto}
\end{figure}

While Fig.~\ref{fig:slitave} shows the slit average of the 2-D spectra
in Figs.~\ref{fig:mu01} and \ref{fig:mu05}, Fig.~\ref{fig:pshisto}
shows the histogram distributions of all the pixels along the slit of
the same Stokes spectra, for $Q/I$ (solid curves) and $U/I$ (dashed
curves) of the 5250.22\,\AA\ line. The quantity that is represented by
the histograms is the amplitude difference between the $\pi$ component
and the average of the two $\sigma$ components, which is insensitive
to errors in the zero point of the polarization scale or the continuum level. 

While the $U/I$ distributions are symmetric around the zero point for
both $\mu$ positions as expected for symmetry reasons, the $Q/I$
distributions are strongly asymmetric, showing that the angular
distributions differ significantly from the isotropic, symmetric case. For
$\mu=0.1$, the asymmetry strongly favors negative values,
implying preferentially horizontal fields, while
it strongly favors the positive values for $\mu=0.5$, implying 
preference for vertical fields. Since the widths of the distributions
are substantially larger than the 1-$\sigma$ error bars, the spread of
the values is of solar origin for the most part (as seen with the
resolving power of the telescope). 

One may think of the distributions in Fig.~\ref{fig:pshisto} as one
part of a sum that is symmetric and the rest that is
non-symmetric. When forming the slit averages to 
obtain Fig.~\ref{fig:slitave}, the contribution from the symmetric part
vanishes, and all the contribution comes from the lop-sided
non-symmetric part. 

While the symmetry properties of the histograms are those of
the angular scaling functions $g_Q$ and $g_U$ that were discussed in
Sect.~\ref{sec:refframe}, the histogram shapes cannot be directly
compared with these functions, since the observed $Q/I$ and $U/I$ are
proportional not only to the angular functions but also approximately
to $B^2$, the square of the field strength. In addition, the shape of the
angular distribution is a strong function of field strength. The
importance of the angular scaling functions of
Sect.~\ref{sec:refframe} is to show how the preference for a
horizontal or a vertical field distribution arises from the sign of
the $g_Q$ asymmetry, independent of any assumption for the $B$
distribution.

\subsection{Height transition from vertical to horizontal
  preference}\label{sec:transition}
So far, we have focused our attention only on the two recordings
represented by Figs.~\ref{fig:mu01} and \ref{fig:mu05}, to illustrate 
very explicitly the physical basis for our conclusions about the
angular distributions. In our next figures, we represent diagrams of
the types in Figs.~\ref{fig:slitave} and \ref{fig:pshisto} by a
single data point per $\mu$ value and spectral line. 

From our 5247-5250 data set of 14 $\mu$ positions, seven represent
regions devoid of any faculae, which we select here as representative of
the quiet-sun background (or intranetwork) fields. We have verified that
there is no subtle influence from any facular point by comparing the results
for two separate slit averages: (1) taking the average of all the
pixels, or (2) only averaging the pixels for which $V/I <
0.5$\,\%. For all the seven selected regions, the two slit averages
result in the same $Q/I$ profile amplitudes, in clear contrast to the
remaining seven regions with faculae. For the facular regions
the full slit averages always show the fields to be strongly
vertically oriented at all $\mu$. There is only evidence of horizontal fields near
the limb for the weak background fields. 

The seven background-field regions consist of the five recordings along the
central meridian from the heliographic N pole from $\mu=0.1$ to 0.5 in
steps of 0.1 plus two recordings near the heliographic E limb at
$\mu=0.08$ and 0.132. 

So far, we have focused the discussion on the behavior of the
5250.22\,\AA\ line, since it is the most Zeeman-sensitive of the lines
and therefore has the highest S/N ratio. We have not yet made use of
the opportunity of having a companion line
(5247.06\,\AA) within the field of view with the same line-formation
properties that only 
differ in Land\'e\ factor. For the transverse Zeeman effect, the
expected linear polarization amplitudes of the two lines should (in
the weak-field limit, for $B\la 500$\,G) be approximately 
proportional to the square of the respective Land\'e\ factors for the
5250/5247 ratio, thus proportional to $9/4=2.25$. We should 
expect to find the same results from the two lines but with a smaller
amplitude in the 5247\,\AA\ line. 

\begin{figure}
\resizebox{\hsize}{!}{\includegraphics{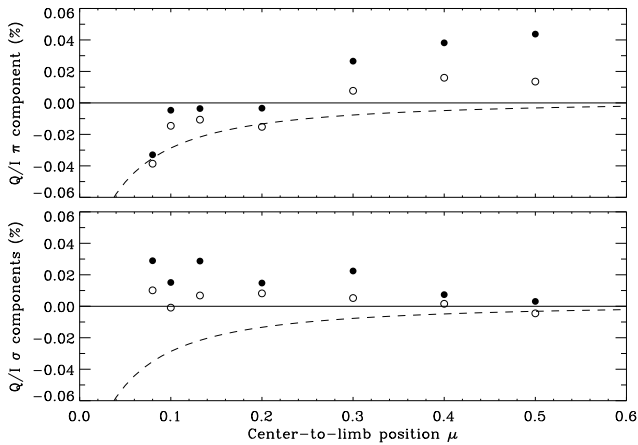}}
\caption{Amplitude of the $Q/I$ $\pi$ component (upper panel) and the
  average of the blue and red lobe $\sigma$ components (bottom panel)
  as a function of $\mu$ for the 7 solar regions with only
  weak background intranetwork-type fields. Filled circles:
  5250.22\,\AA\ line. Open circles: 5247.06\,\AA\ line. The dashed
  line represents the position of the polarization zero level before
  being shifted to the continuum level. 
}\label{fig:qsigpi}
\end{figure}

In Fig.~\ref{fig:qsigpi}, we have plotted the $\pi$ component
amplitudes that have been extracted from the seven recordings (upper
panel) and the average of the blue and red $\sigma$ components (bottom
panel) as filled circles for the 5250\,\AA\ line, as open circles for
the 5247\,\AA\ line as a function of $\mu$. For reference, the dashed
line represents the true position of the polarization zero level owing to the continuum
polarization before shifting the zero to the continuum level used as
the reference level for the Zeeman effect analysis. 

In Fig.~\ref{fig:qsigpi}, we have not plotted the $\pi -\sigma$
difference like in Fig.~\ref{fig:pshisto} but plotted the absolute values of
the separate components. These values are susceptible to errors in the
polarization zero level used, and the errors may, in addition, be different for
the two lines, if they are caused by weakly polarized fringes. This is the reason
why the positions of the points look inconsistent and
demonstrates why it is so important to use differential measures, such
as the $\pi -\sigma$ amplitude difference used in Fig.~\ref{fig:pshisto}. This
differential measure is obtained by subtracting the values in the
lower panel of Fig.~\ref{fig:qsigpi} from the corresponding values in
the upper panel. It is not meaningful to go into a detailed
interpretation of Fig.~\ref{fig:qsigpi} before having formed these 
differential measures. Instead, the figure illustrates the possible
pitfalls of an analysis that is not differential. 

\begin{figure}
\resizebox{\hsize}{!}{\includegraphics{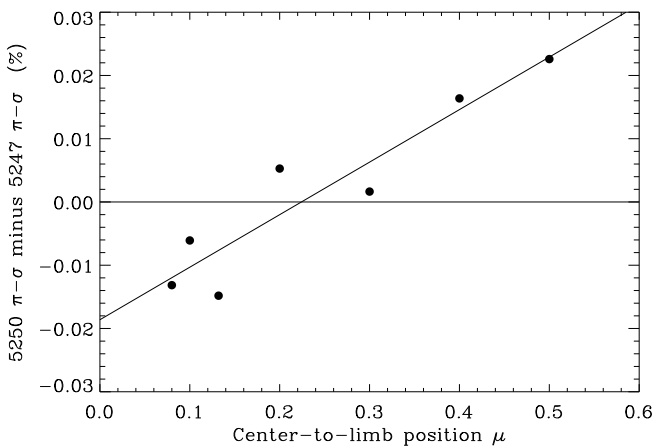}}
\caption{Difference between the $\pi$ and $\sigma$ components for the
  5250\,\AA\ line minus the corresponding difference for the
  5247\,\AA\ line, plotted as a function of $\mu$ for our seven 
  intranetwork regions. The negative values show that the angular
  distribution is preferentially horizontal limbwards of  $\mu=0.22$ (the zero crossing of the straight-line fit), while it is preferentially vertical on the diskward  side. 
}\label{fig:pisig}
\end{figure}

These $\pi -\sigma$ differentials can be formed separately for the 5250
and 5247\,\AA\ lines from the difference between the two panels in
Fig.~\ref{fig:qsigpi}. These differentials should show the same
behavior for the two lines, except with much smaller amplitude for the
5247\,\AA\ line. In Fig.~\ref{fig:pisig}, we go a step further to
create a double differential by forming the difference between the
$\pi -\sigma$ differential for the 5250\,\AA\ line and the
corresponding differential for the 5247\,\AA\ line. By doing so, we
doubly suppress systematic errors that can arise from fringes and varying
polarization background levels. 

Besides random measurement noise the remaining scatter of the points
around the straight-line fit in Fig.~\ref{fig:pisig} is mainly related to 
the finiteness of our modest statistical sample of solar resolution
elements (140 effective spatial pixels for each of the 7 Stokes
images). In this sense, the scatter of the points is partly solar 
in origin, and it is therefore not so meaningful to assign error bars to
them. Note that the standard deviation of the scatter of the points around
the straight-line fit is only $5\times 10^{-5}$, much of which is of
solar origin. 

In terms of this doubly differential measure, positive values
correspond to angular distributions that are more peaked around the
vertical direction (in comparison with the isotropic distribution),
while negative values represent angular distributions that favor the
horizontal plane. The zero crossing of the straight-line fit in
Fig.~\ref{fig:pisig} occurs at $\mu=0.22$. Limbwards of this
center-to-limb position, the intranetwork fields are preferentially
horizontal, while diskwards they are preferentially vertical. This
confirms the qualitative conclusions in a more quantitative way that we
could draw from plain inspection of the slit-averaged Stokes spectra.

\section{Interpretation and conclusions}\label{sec:interpret}

\subsection{Height and flux-density dependence}\label{sec:height}
The height of line formation increases with decreasing $\mu$, when we
move closer to the limb. Our finding that the $\pi -\sigma$ 
amplitude changes sign around $\mu=0.22$ for the weak background
fields implies that the angular distribution of the field vectors is
more pancake-like (horizontal) in the upper photosphere near the
temperature minimum and above, while it is more vertical
in the middle photosphere and below. For disk-center observations of
spectral lines like the 5247-5250 or the 6301-6302 line pairs, the
distribution is always more vertical (than the isotropic case), which
confirms our previous conclusions from the analysis of Hinode observations
\citep{stenflo-s10aa}. 

There is a close relation between angular distribution and 
flux density. The stronger flux densities remain peaked around
the vertical direction for the whole height range covered by our
various $\mu$ values. This is consistent with our previous
Hinode results \citep{stenflo-s10aa} that the degree of
concentration around the vertical direction increases steeply with
flux density, while the distribution tends to become nearly isotropic
in the limit of low flux densities. In the present work, we find 
that as we move up in the atmosphere, the distribution does not approach the
isotropic case in the limit of low flux densities but changes from
being more vertical to becoming more horizontal above a certain level. 

Here, we have made the distinction between solar regions containing
faculae and regions with background fields devoid of facular points,
but we could also have used the terminology network and
intranetwork. Thus network flux remains preferentially vertical
throughout the considered height range. In contrast, intranetwork fields are
preferentailly vertical in the lower and middle photosphere but become preferentially
horizontal in the higher layers. 

While there is a correspondence between $\mu$ value and geometrical height,
a quantitative translation of $\mu$ into a height scale would require
radiative-transfer modelling, which is outside the scope of the
present work.

\subsection{Roles of buoyancy and stratification}\label{sec:buoy}
The outer envelope of the Sun is highly stratified with a steep
outwards decrease in the gas pressure. This has important
consequences for the height dependence on the magnetic field, which is
governed by three main factors: buoyancy, containment of the magnetic
pressure, and topology (in particular the scales of polarity mixing). 

As the magnetic flux concentrations are partially evacuated to satisfy pressure
equilibrium with the surroundings, they become subject to strong buoyancy
forces. Field lines that are anchored (by the frozen-in condition) in
the dense deeper layers will be forced by buoyancy towards the 
upright orientation. Buoyancy tries to generate angular
distributions that are peaked around the vertical direction. 

The equilibrium between magnetic pressure and ambient gas
pressure shifts rapidly in favor of the magnetic pressure as we move
up in height. This means that any magnetic-field concentration will
expand and the field lines flare out. For this reason, the angular
distribution will contain a larger 
proportion of highly inclined fields in the higher layers. 

Numerical simulations of magneto-convection do indeed show that
  the magnetic field becomes more horizontal in the upper photospheric
layers
\citep{stenflo-abbett07,stenflo-schvoeg08,stenflo-steiner_etal08}.
The maximum height of magnetic flux loops scales with the separation 
of their footpoints at the bottom of the photosphere. Near the top of the
loops, the flux is largely horizontal. The smaller the polarity mixing
scale of the flux footpoints, the lower is the loop height, which
means a lower transition height in the atmosphere from a vertical
to a horizontal field distribution \citep[cf.][]{stenflo-steiner10}. 

We thus have a competition between buoyancy, which favors vertical
fields, and the other two effects (flux expansion and polarity
mixing), which lead to an increasing proportion of 
inclined fields as we move up in height. We expect the height at
which buoyancy loses to the other effects 
to depend on field strength, since the magnitude of the
buoyancy force scales with field strength. Therefore,  
stronger fields are expected to retain their vertical preference at
greater heights, as
observed. In contrast, the weakest fluxes are not only less affected
by buoyancy, but they are also more easily buffeted by the
dynamic pressure of convective turbulence and thereby develop more small-scale polarity
mixing. While the quiet-sun network fields tend to be fairly unipolar
(on the supergranulation scale), the intranetwork fields are thus 
characterized by mixed polarities on small scales. 

Observations with 3\,arcsec spatial resolution of the temporal
  variations in the line-of-sight 
  component of this intranetwork
  field reveal that it is highly
  dynamic and that its fluctuations get much more
  pronounced near the solar limb \citep{stenflo-harvey_etal07}. This
  is evidence of a dynamic horizontal field component in the upper
  photosphere (where the spectrum observed 
  near the limb originates). A related result has been obtained from
  Hinode observations by
  \citet{stenflo-lites11ssd}, who finds an increased line-of-sight flux density
  for the intranetwork fields as one goes from disk center towards the
  solar limb. Part of this effect is owing to the circumstance that the
  intranetwork field has a wide angular distribution, and part of it
  is a height variation. However, these observations do not tell in which sense the
  angular distribution deviates from the isotropic case (for instance
  in favor of horizontal fields), and whether this deviation changes
  sign with height. This is the issue that is being
  addressed by our present analysis.

Hanle-effect observations have shown that the photosphere is seething
with an ocean of 
mixed-polarity fields on scales on the order of a few km or less. The
  atmospheric stratification becomes increasingly irrelevant as we go
  to smaller and smaller tangled magnetic structures in the size
  domain below the scale height. Because of 
  scale separation, the angular distribution may be expected to
  approach the isotropic case in the small scale limit.

\subsection{Model independence of the conclusions}\label{sec:modelind}
Our method to determine whether the angular distribution of field vectors
is more horizontal or more vertical as compared to the isotropic
distribution is model independent in the sense that it only depends on
the fundamental symmetry property of the transverse Zeeman effect: If
$\pi -\sigma$ for the spatially averaged $Q/I$ is positive, then the field distribution
is more vertical, while it is more horizontal if it is negative. The
magnitudes of the polarization values never enter in this decision,
only the sign. By forming double differentials, like the difference of
the $\pi -\sigma$ values between the 5250 and the 5247\,\AA\ lines, we suppress
possible systematic errors. 

This approach does not depend on the particular
choice of parametrization of the magnetic-field distribution functions
given by Eq.~(\ref{eq:a}), since the qualitative symmetry properties
describing the relation between the sign of the average $Q/I$ and the
horizontal-vertical preference would be the same with any choice. 

Note that we do not try in the present paper to convert polarization
values into field strength or $\mu$ values into geometrical height 
but limit our determination to the variation in the sign of the
vertical-horizontal preference with center-to-limb distance. A
conversion to geometrical height and to field strength values at these heights
would require radiative-transfer modelling and make the results model
dependent. By letting this be outside the scope of the present paper, 
we keep the results that are presented here model independent in the sense described.

\subsection{Independence of angular resolution}\label{sec:resolution}
While improvements of angular resolution are necessary to advance our
knowledge about the morphology and evolution of 
solar magnetic fields, several of the main fundamental insights into
the nature of quiet-sun 
magnetic fields have not depended on the resolving power of the 
telescope used. Examples are as follows: The extreme intermittency
with much of the total quiet-sun 
magnetic flux being carried by collapsed, kG-type fields was
discovered with an angular 
resolution of several arcsec thanks to the near model-independence
that is a unique property of the 5250-5247 line ratio 
\citep{stenflo-s73}. The discovery that the photosphere is seething 
with vast amounts of hidden magnetic flux with strengths in the
range of 10-100\,G was made by interpreting observations of the Hanle
effect depolarization with a resolution on the order of one arcmin 
\citep{stenflo-s82}. Similarly, the use of distribution
functions for the transverse Zeeman effect to derive model-independent
constraints on the angular distribution functions of quiet-sun
magnetic fields, which has been the topic of the present paper, 
does not depend on the resolving power of the telescope, as
  clarified in Sect.~\ref{sec:telres}. Although this
method was known and applied more than a quarter of a century ago
\citep{stenflo-s87}, it is only now that we are in a position to more
fully exploit it. The reason is not because of the improved angular
resolution but because of the advances in high-precision imaging Stokes
polarimetry, in particular through the availability of the ZIMPOL
technology. In contrast, the 1987 investigation was based on observations with
single-pixel detectors (photomultipliers).

\begin{acknowledgements}
I am grateful to the THEMIS Director, Bernard Gelly, for making this
powerful telescope available to us for our ZIMPOL observing
campaigns. The ZIMPOL team that participated with the author in the 2008 campaign 
consisted of Daniel Gisler and Lucia Kleint (ETH Zurich), Renzo Ramelli (IRSOL),
and Jean Arnaud (Toulouse). ZIMPOL was invented and developed by
Hanspeter Povel (ETH Zurich) with long-term support at ETH by Peter Steiner and
Frieder Aebersold. 
\end{acknowledgements}



\end{document}